\documentclass[twocolumn]{aastex701}
\usepackage{subcaption}
\usepackage{xfrac}
\begin{document}

\title{An Investigation of 5-year Simultaneous X-ray and Radio Light Curves of the Dwarf Seyfert Galaxy UGC 6728}

\author[0000-0001-5785-7038]{Krista Lynne Smith}
\affiliation{George P. and Cynthia Woods Mitchell Institute for Fundamental Physics and Astronomy\\ Texas A\&M University\\ College Station, TX 77843-4242, USA}
\email[show]{kristalynnesmith@tamu.edu}

\author[0000-0002-1292-1451]{Macon Magno}
\affiliation{George P. and Cynthia Woods Mitchell Institute for Fundamental Physics and Astronomy\\ Texas A\&M University\\ College Station, TX 77843-4242, USA}
\affiliation{CSIRO Space and Astronomy, ATNF, PO Box 1130, Bentley WA 6102, Australia}
\email[show]{macon.a.magno@tamu.edu}

\author[0000-0002-7998-9581]{Michael Koss}
\affiliation{Eureka Scientific, Inc., 2452 Delmer Street, Suite 100, Oakland, CA 94602-3017, USA}
\email{m ike.koss@eurekasci.com} 

\begin{abstract}

We present serendipitous simultaneous radio and X-ray light curves of the dwarf Seyfert galaxy UGC~6728 spanning 5 years. The X-ray light curve exhibits a flaring period, followed by a gradual rise and decline. Throughout these events, the X-ray hardness ratio and spectrum do not change significantly. The radio flux is constant, as far as can be determined from its sparse sampling, until the end of the X-ray flare, then decreases by a factor of two by the midpoint of the gradual X-ray rise before returning to baseline at the end of the X-ray decline. We interpret this behavior in light of a similar event recently reported in NGC~2992, in which there is a temporary obscuration of the radio source by a blob of plasma ejected by a magnetic reconnection in the accretion disk. The energetics of the X-ray flare are consistent with those expected from magnetic disk activity. As in NGC~2992, the X-ray spectrum does not evolve during the obscuration event. We also discuss the possibility that the observed phenomena are due to normal AGN coronal flaring and variability, which is plausible but unlikely given the lack of spectral variation.
\end{abstract}


\section{Introduction} 

Radio-quiet active galactic nuclei (AGN) are the energy sources of the large majority of active galaxies in the local universe. The geometry and causal relationships in the central engines of these objects are not directly observable, and must be inferred from the apparent relationships between different wavebands based on a few important assumptions: that the optical and UV emission is primarily thermal, from an accretion disk or flow \citep{Shields1978,Malkan1982}; the X-ray emission is due to inverse Compton scattering of the disk photons by a central plasma known as the ``corona" \citep[e.g., ][]{Haardt1993}; and the radio emission may be due to synchrotron emission from that same coronal plasma \citep{Laor2008} or from a small, weak analog of the powerful radio jets seen in the less-common radio-loud AGN \citep[][along with other interpretations]{Panessa2019} with some evidence that its origin may depend on Eddington ratio \citep{Alhosani2022}. Simultaneous timing across wavebands offers an especially revelatory window into the interrelationships of the accretion disk and the X-ray emitting coronal plasma, exemplified in the exquisite results of the AGN STORM collaboration \citep[e.g., ][]{Edelson2015,Kara2021,Cackett2023}. 
Frequently missing from such efforts is simultaneous monitoring in the radio, due mainly to challenging observational logistics. However, such monitoring is necessary to pin down the radio emission's origin; for example, correlated variability with minimal lag between X-rays and radio may be expected if they share a coronal origin, while a lag may be expected if the radio emission is at least partially due to a jet coupled to the disk-corona system.

Although relatively rare, there have been a handful of simultaneous monitoring campaigns of radio-quiet AGN in X-rays and radio. In one of the earliest, \citet{Bell2011} found a weak correlation with a 20-day time lag between the X-ray and radio in the low-luminosity radio-intermediate AGN NGC~7213, and interpreted this in light of the predicted disk-jet coupling from the fundamental plane of black hole activity \citep{Merloni2003}. More recently, \citet{Panessa2022} reported rapid variability in the 5~GHz radio and soft X-ray emission of Mrk~110, placing stringent limits on the size of the radio emission region of $\sim$180 Schwarzschild radii and did not detect any correlation between variability in the two wavebands. \citet{Chen2022} presented overlapping archival light curves from RXTE and the VLA of three radio-quiet Seyferts with diverse radio spectral properties and found a few tentative time-delayed correlations, but none with statistical significance. A multi-wavelength campaign of the changing-look AGN 1ES~1927+654, which apparently began a major state change in 2017 which persisted at least until 2023, and has been explained as a possible tidal disruption event \citep{Ricci2021}, but is more likely due to change in the accretion flow, possibly leading to an inversion of the magnetic field direction near the horizon \citep{Laha2022}. During this event, the ratio between the X-ray and radio luminosity varied but remained broadly consistent with the coronal relation derived from active stars \citep{Gudel1993}; however, the radio flux does not return to pre-2017 levels, as best can be determined over the four post-flare observed epochs \citep{Ghosh2023}. \citet{Petrucci2023} monitored the well-studied radio-quiet AGN MCG+08-11-11 in the millimeter and X-ray wavebands simultaneously, and found similar increasing trends and rapid variability, suggesting a coronal origin. Finally, of particular relevance to this work, \citet{Fernandez2022} monitored the radio-quiet Seyfert NGC~2992 with \emph{Swift}-XRT and the VLBA over six months and observed a significant decrease in the radio flux density and a simultaneous increase in the X-ray flux, but with no significant change in the X-ray spectrum. They interpret these results in light of a blob of ionized material, ejected from the accretion disk during a period of increased X-ray activity, and rising to temporarily increase the opacity at radio frequencies along the line of sight, causing the radio dimming. 

A few studies have compared the very high-frequency radio or mm-band variability to X-rays in samples of objects, rather than individual sources, with mixed results. The consensus is that the statistical properties of the variability between the two bands are much the same, implying that they share a physical origin - usually postulated as the X-ray emitting coronal plasma \citep{Baldi2015,Behar2020}. Such an interpretation is supported by the strong correlation between the 100~GHz and X-ray luminosity in large samples \citep{Ricci2023}.

\begin{figure*}[ht]
\gridline{\fig{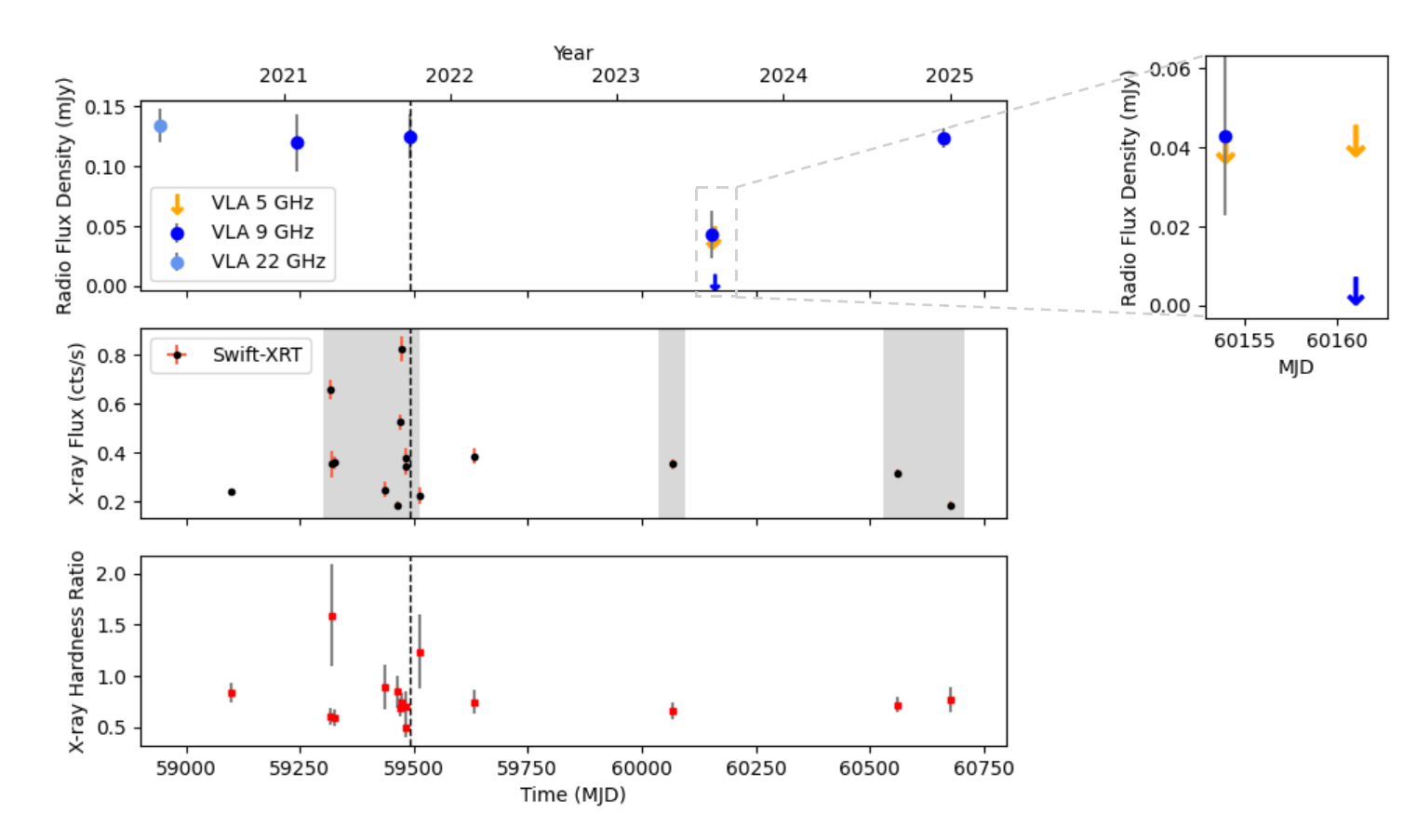}{\textwidth}{}}
\vspace{-1cm}
\caption{Radio, X-ray, and X-ray hardness ratio curves. Each point includes a 1$\sigma$ error bar, which are sometimes comparable to the marker size. A black dashed line is provided across all panels at the position of the first post-flare radio data point for ease of comparison. The region of the radio light curve where the C-band upper limits were observed is blown up on the right. These observations were simultaneous with the X-band observations in the same program. Both bands detected only upper limits on MJD 60161. Grey regions in the X-ray light curves represent the flare, middle, and end spectral extraction regions as described in Section~\ref{sec:xray_data}.}
\label{fig:lcs}
\end{figure*}

In this work, we report on the serendipitous archival simultaneous light curves of the radio-quiet Seyfert UGC~6728, a particularly low-mass Seyfert galaxy (or ``dwarf Seyfert") with a black hole mass of only $M_\mathrm{BH} = 7.1\pm{0.4} \times 10^{5} M_\odot$ in a barred lenticular galaxy with a low stellar mass of log $M_* / M_\odot = 9.9\pm0.2$ at a redshift of 0.0065\footnote{Note: there are conflicting redshift values given in SIMBAD and NED for this object. We choose the value given by NED, $z=0.0065$, which is used in previous studies on UGC~6728 and which yields a luminosity distance of 28.1~Mpc assuming the cosmological parameters of \citet{Bennett2014}.}\citep{Bentz2016,Bentz2021,Nandi2024}. It is also thought to be a ``bare" AGN, with an extremely low hydrogen column density along the line of sight to the nucleus \citep{Walton2013,Nandi2024}.
Although the light curves are poorly sampled, it has the same number of cadences as the NGC~2992 event and exhibits similar behavior, with the addition of a prominent X-ray flare preceding the possible obscuration event, consistent with expectations for the magnetically-released blob model. We investigate these data in light of the scenario presented for NGC~2992 by \citet{Fernandez2022}, to determine whether it is a feasible candidate for the same phenomenon and briefly discuss other possible interpretations. The paper is organized as follows. In Section~\ref{sec:data} we discuss the provenance and reduction of the radio and X-ray archival data. In Section~\ref{sec:lc_results} we present the final light curves, which we interpret in light of the model in Section~\ref{sec:discussion}. We summarize our results briefly in Section~\ref{sec:conclusion}.

\section{Observations and Analysis}
\label{sec:data}
\subsection{Radio Data and Analysis}
\label{sec:radio_data}

We discovered UGC 6728’s unusual behavior serendipitously during a widespread search for archival radio observations of radio-quiet Seyferts in the 70-month \emph{Swift}-BAT catalog \citep{Baumgartner2013}. There are five archival X-band radio observations of UGC~6728, obtained in programs 20B-211 (PI O'Dea), SH0126 (PI Donahue), 23A-349 (PI Daly), and 24B-387 (PI Koss). Four were taken with the array in A-configuration with beam major axes ranging from $\sim0.29 - 0.44$~arcsec; one (SH0126) was taken with the array in B-configuration with a beam major axis of 0.96~arcsec.  There are also two archival C-band observations in program 23A-349, simultaneous with the X-band data. For all of these, we processed the raw data through the standard VLA reduction pipeline using Common Astronomy Software Applications (CASA) \citet{Casa2022} v.6.5.4. For each archival observation, we split off UGC~6728 from the measurement set and cleaned the observation with the \verb|tclean| task in CASA to a threshold of 0.05 mJy with Briggs weighting (\verb|robust| = 0.5).  No other potential contaminating sources were present in the 1024\arcsec~field to complicate cleaning. The source was unresolved in all images, with predicted source dimensions smaller than the beam size. There is no indication of extended emission, so we do not consider the beam size problematic in the flux of the one B-array observation. In any case, the flux density of this observation is not higher than the previous one (taken in the A-array). We then extracted the flux density of the point source for each image using the beam-fitting capability of  Cube Analysis and Rendering Tool for Astronomy (CARTA) \citet{Comrie2024}. We also include in the following analysis the K-band observation included \citet{Magno2025}, from program 20A-158 (PI Smith). The radio flux measurements, rms noise values, and beam properties are given in Table~\ref{t:radio_lc}. Both archival C-band observations were non-detections, as was the fourth X-band observation (simultaneous with the second C-band one). As shown in Table~\ref{t:radio_lc}, the noise properties of the non-detection images are similar to those of detections, so the non-detections are not a result of a higher noise floor.

\begin{figure*}[t!]

    \begin{subfigure}[t]{0.37\textwidth}
        
        \includegraphics[height=2.7in]{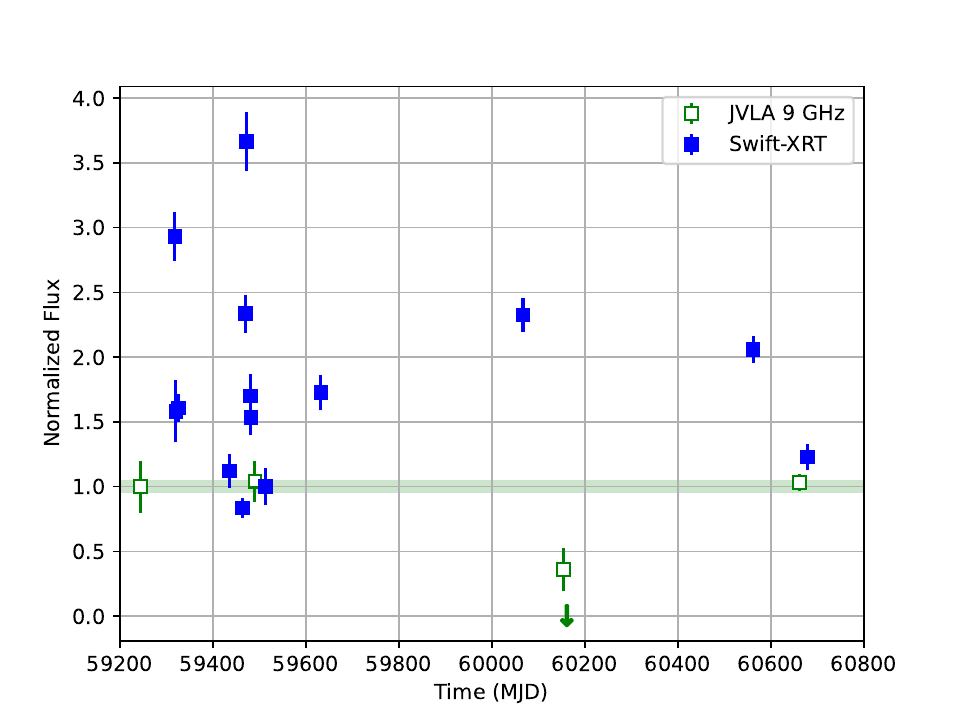}
        \caption{}
    \end{subfigure}%
    \hspace{2cm}
    \begin{subfigure}[t]{0.37\textwidth}
        \centering
        \includegraphics[height=2.7in]{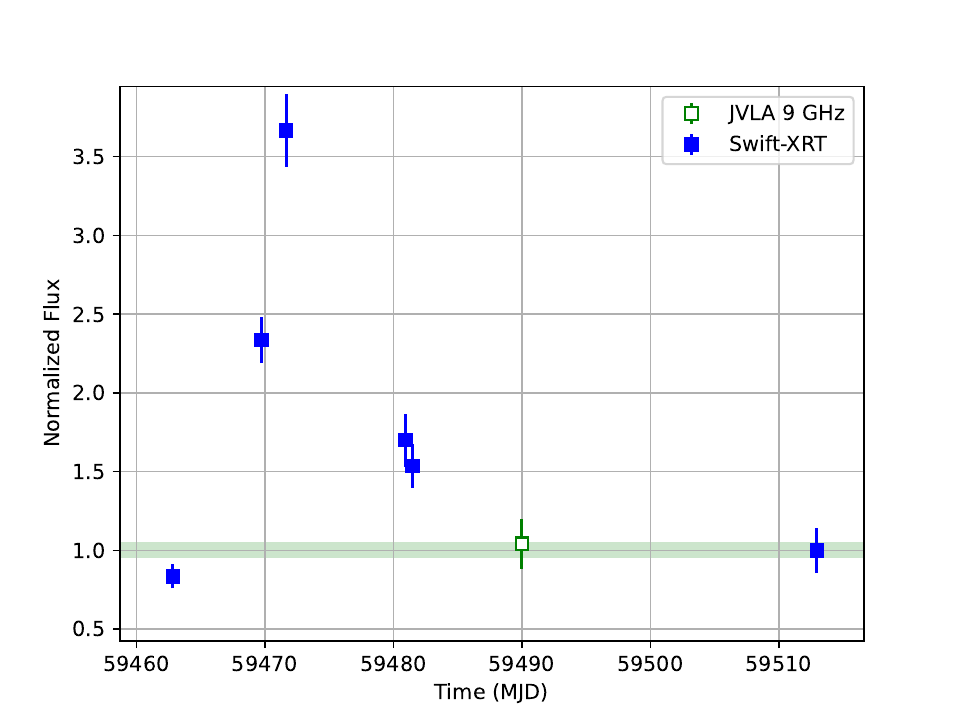}
        \caption{}
    \end{subfigure}
    \caption{\emph{Left: } Normalized light curves, with radio fluxes normalized to the first radio data point and X-ray normalized to the first X-ray point after the flare. \emph{Right: } Same as first panel, but zoomed in on the X-ray flare. Colors and normalizations are chosen to match those presented for NGC~2992 by \citet{Fernandez2022} to facilitate comparison with their Figure~11.}
\label{fig:normalized_lcs}
\end{figure*}

We present the full radio light curve in the top panel of Figure~\ref{fig:lcs}, and again in the normalized light curve for comparison to the X-rays in Figure~\ref{fig:normalized_lcs}.

\subsection{X-ray Data and Analysis}
\label{sec:xray_data}
The \emph{Swift}-XRT data were obtained from the High Energy Astrophysics Science Archive Research Center (HEASARC) for the period relevant to the radio observations: all available PC-mode observations since MJD~59244 (January 30, 2021). UGC 6728 was observed in two separate target of opportunity requests (PI Koss) concurrent with planned \emph{Hubble} Space Telescope (HST) observations and the final radio data point (also PI Koss).  These products were compiled using the UK Swift Science Data Centre online service \citep{Evans2009}. For flux analysis, the count rates in the light curve were transformed into fluxes using PIMMS v.4.15 \citep{Mukai1993} assuming a power law spectral model with the spectral index calculated in the XSPEC fitting for that light curve epoch, as described immediately below. Observation IDs, count rates, and fluxes are given in Table~\ref{t:xray_lc}. The hardness ratio is calculated using a soft band of 0.3-1.5~keV and a hard band of 1.5-10\,keV and constructing the simple ratio of $H/S$. 

\begin{figure}
\gridline{\fig{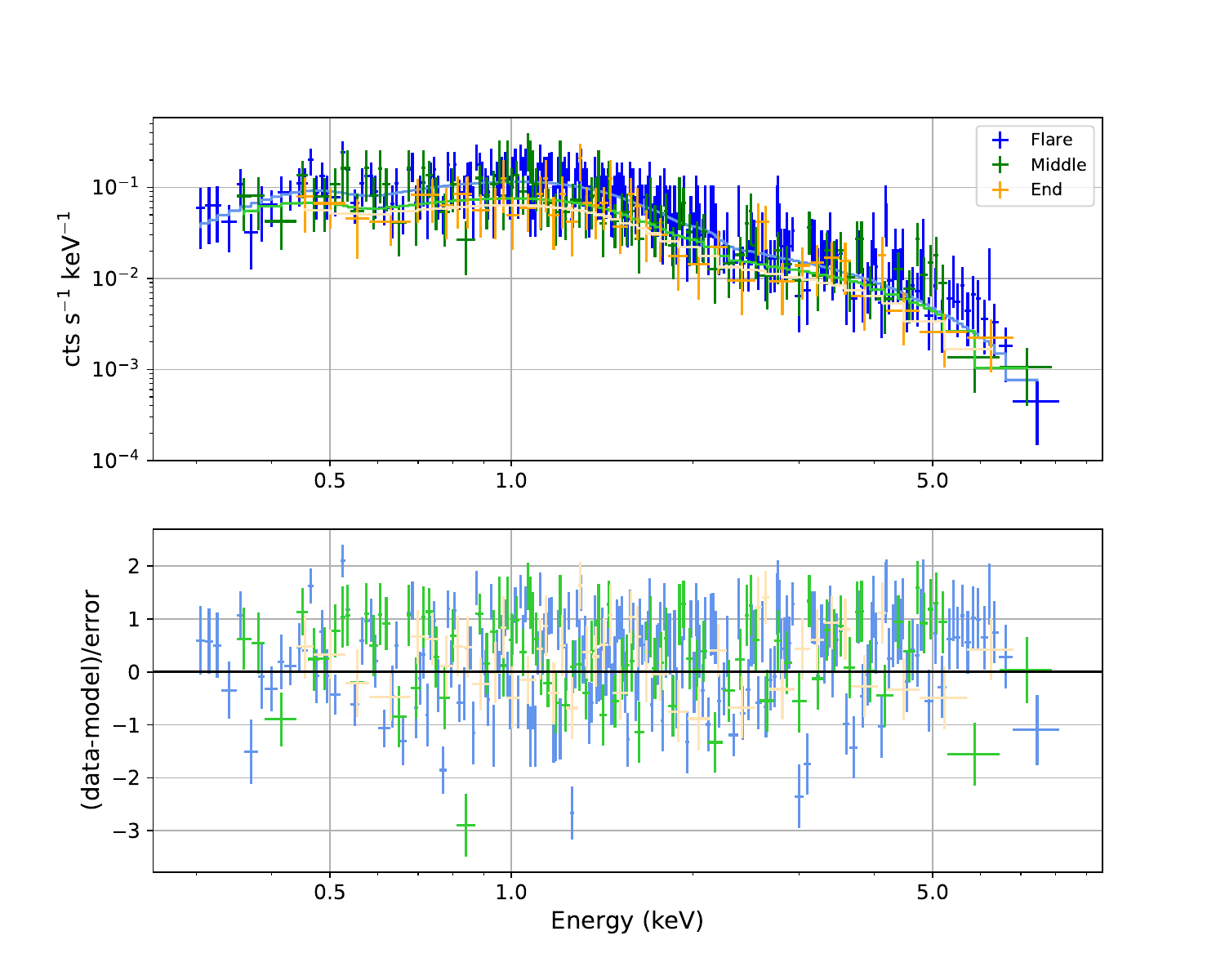}{0.5\textwidth}{}}
\caption{X-ray spectra of the X-ray flare and the middle and end of the radio-dimming event, as defined in Section~\ref{sec:xray_data}. The best-fitting model for each spectrum, from the parameters given in Table~\ref{t:xrayspec}, is shown as a solid step-line. The lower panel shows the fit residuals.}
\label{fig:xray_spectra}
\end{figure}

The light curve was separated into three regions for spectral analysis: all epochs up until 59512, which we deem the ``flare" period; the observation in the middle of the radio decrease, which we call the ``middle" period; and the two observations at the end when the X-ray flux decreases, which we call the ``end" period. These periods are shown in the grey shaded regions on Figure~\ref{fig:lcs}. For each period, we first bin the spectra using \texttt{grppha} to obtain a minimum of 3 counts per bin. We then fit the spectrum using \texttt{XSPEC} \citep{Arnaud1996}, ignoring all channels with $E<0.3$~keV and $E>7$~keV and all spectral bins flagged as bad. We use a \texttt{phabs * zphabs * zpow} model, as the spectral quality is not sufficient for more detailed physical models, and following the method of \citet{Fernandez2022}. The Galactic absorption in the \texttt{phabs} term is set at $4.4\times10^{20}$cm$^{-2}$, based on the maps from \citet{HI4PI2016}. The results of the spectral fitting are listed in Table~\ref{t:xrayspec} and the spectra for each region are shown together in Figure~\ref{fig:xray_spectra}. For the latter two epochs, the intrinsic column density in the source is consistent with only Galactic extinction, as expected for an object considered a ``bare" AGN \citep{Walton2013}.

\begin{table}[t]
\textbf{Radio Light Curve Data}
\centering
\begin{tabular}{||c c c c c||} 
 \hline
Program & MJD & $S_\nu$ & Noise rms & Beam Size \\ [0.5ex] 

   & & mJy & $\sfrac{\mu \mathrm{Jy}}{\mathrm{beam}}$  & arcsec \\
 \hline \hline
C-Band: & & & &\\
\hline
 23A-349 & 60154 & $<0.0040$ & 13.4 & $0.49\times0.31$ \\
  23A-349 & 60161 & $<0.0042$ & 13.9 & $0.54\times0.29$ \\
 \hline\hline

X-Band: & & & & \\
\hline
20B-211 & 59244 & $0.120\pm 0.024$ & 13.6 & $0.29\times0.18$  \\ 
SH0126 & 59490 & $0.125\pm0.019$ &  10.3 & $0.97\times0.60$ \\
23A-349 & 60154  & $0.043\pm 0.02$  &  12.6 & $0.30\times0.21$\\
23A-349 & 60161 & $< 0.065$ & 12.8 & $0.35\times0.18$ \\
24B-387 & 60661 &$0.124 \pm 0.008$ & 4.7 & $0.44\times0.17$ \\
\hline \hline
K-Band: &&&& \\
\hline
20A-158 & 58944 & $0.134\pm0.014$ &  10.6 & $1.29\times0.79$ \\
\hline
\end{tabular}
 \caption{Radio program IDs, epochs, flux density values, and beam dimensions. The observation in SH0126 was taken in B-array and the observation in 20A-158 was taken in C-array; all others were taken in A-array.}
 \label{t:radio_lc}
\end{table}

\begin{table}
\textbf{X-ray Light Curve Data}
\centering
\begin{tabular}{||c c c c c||} 
 \hline
OBSID & MJD & Ct Rate & Flux & HR \\ [0.5ex] 

    &  & $e^-$ s$^{-1}$  &erg cm$^2$ s$^{-1}$ & \\
 \hline\hline
00013662006 &	59097	&	0.240	&	4.21E-12	&	$	0.85	\pm	0.10	$	\\
00013662007 & 	59317	&	0.657	&	1.15E-11	&	$	0.60	\pm	0.08	$	\\
00035266001 &	59320	&	0.353	&	6.21E-12	&	$	1.59	\pm	0.50	$	\\
00035266002 &	59325	&	0.360	&	6.31E-12	&	$	0.60	\pm	0.09	$	\\
00035266003 &	59435	&	0.251	&	4.40E-12	&	$	0.89	\pm	0.22	$	\\
00081098001 & 	59462	&	0.187	&	3.28E-12	&	$	0.85	\pm	0.16	$	\\
00088256001 &	59469	&	0.523	&	9.17E-12	&	$	0.69	\pm	0.09	$	\\
00096132002 &	59471	&	0.821	&	1.44E-11	&	$	0.75	\pm	0.10	$	\\
00096132003 &	59480	&	0.381	&	6.68E-12	&	$	0.71	\pm	0.14	$	\\
00096132004 & 	59481	&	0.344	&	6.03E-12	&	$	0.50	\pm	0.09	$	\\
00096132005 & 	59512	&	0.224	&	3.93E-12	&	$	1.24	\pm	0.36	$	\\
00096132006 &	59631	&	0.387	&	6.79E-12  &	$	0.75	\pm	0.12	$	\\
00096132007 &	60066	&	0.354	&	9.13E-12	&	$	0.66	\pm	0.08	$	\\
00096132008 &	60561	&	0.314	&	8.09E-12	&	$	0.72	\pm	0.08	$	\\
00096132011 & 	60678	&	0.187	&	4.82E-12	&	$	0.77	\pm	0.13	$	\\
 \hline

\end{tabular}
 \caption{X-ray observation IDs, epochs, count rates, flux values, and hardness ratios, all from \emph{Swift}-XRT.}
 \label{t:xray_lc}
\end{table}

\section{X-Ray and Radio Light Curves} 
\label{sec:lc_results}

Both light curves show significant variability over the baseline considered ($\sim4.4$~years). The X-ray exhibits a flare from a baseline of $\sim0.2$~cts/s to a maximum of $\sim0.8$~cts/s, with an exponential rise and fall and then returning to the pre-flare baseline after anywhere between 20 and 50 days flare duration. The X-ray flux then gradually rises to about a factor of 2 above the pre-flare baseline of 0.2~cts/s, a level it maintains for about 3 years before returning to baseline. 

There are five radio epochs that correspond to the X-ray light curve described above. One occurs before the X-ray flare, one just as it ends, two near the apex of the X-ray rise (although they are separated by only 5 days), and one near the time when the X-ray returns to baseline. The radio behaves in the opposite manner to the X-rays:  after maintaining a steady flux density of $\sim0.12$~mJy since mid-2020, it decreases by at least a factor of 2 by mid 2023, falling to a minimum of a non-detection with an upper limit of 0.06~mJy and then returning to the original value around the same time as the X-ray falls back to its baseline. However, there are only two observed epochs during the X-ray rise-and-fall event, 5 days apart. The radio behavior is not well sampled during this time, so it is not known whether the factor-of-two dimming in the radio is part of a smooth trend that coincides with the X-ray brightening, or is merely a low point in the normal stochasticity of the AGN (however, the radio flux is nearly constant otherwise). Caution is therefore required in interpretation; we proceed with our comparison to the NGC~2992 event assuming the radio is dimming as the X-ray is rising, and returns to baseline as the X-ray does.

This behavior is the same as that observed in the radio-quiet Seyfert NGC~2992 by \citet{Fernandez2022}, over a similar number of light curve data points but with a significantly shorter baseline for the dimming/rising event of only 140~days. The NGC~2992 data also do not include X-ray monitoring before the event, so in their case no flare could have been observed. 

\begin{table*}[t]
\centering
\begin{tabular}{||c c c c c||} 
 \hline
Spectral Epoch & \texttt{zphabs} $N_H$ & \texttt{zpow} $\Gamma$ & \texttt{zpow} Norm. & Reduced $\chi^2$ \\ [0.5ex] 
 \hline\hline
 Flare & $0.062 \pm 0.024$ & $2.00\pm 0.10$ & $0.0028\pm 0.0002$ & 0.79 \\ 
 \hline
 Middle & - & $1.69\pm 0.12$ & $0.0015\pm 0.0002$ & 0.76 \\
 \hline
 End  & - & $1.70\pm 0.28$ & $0.0010\pm 0.0003$ & 0.48 \\
 \hline

 \hline
\end{tabular}
\caption{Summary of X-ray spectral fitting results for the three epochs defined in Section~\ref{sec:xray_data} and shown by the gray shaded regions in Figure~\ref{fig:lcs}.}
\label{t:xrayspec}
\end{table*}

\section{Results and Discussion}
\label{sec:discussion}

In this section we will discuss the interpretation of the light curves in light of a few scenarios. We primarily consider whether or not these light curves suggest a magnetic reconnection event similar to that proposed to describe the variability seen in NGC~2992, and follow the analysis of \citet{Fernandez2022} for a clear comparison. Given the sparseness of the radio observations, however, definitive support for this interpretation is not possible. Due to this, we also discuss the possibility of a flaring corona creating the observed light curves. Explanations not related to AGN, such as supernovae due to star formation, are not considered as likely origins of radio or X-ray emission in this galaxy, as it has a low star formation rate (even compared to other Seyferts) and a minimal star formation contribution to the mid-infrared emission \citep{Garcia2022,Chen2025}.

\subsection{A Magnetic Reconnection Releasing an Obscuring Plasmoid}
\label{sec:magrecon}

The primary physical model used by \citet{Fernandez2022} to explain the event they observed is a temporary increase in the free-free opacity towards the radio source due to a rising blob of plasma released by the accretion disk. Such plasmoids may be thought of as varying amounts of extended coronal plasma released by magnetic reconnections in the inner disk regions, a scenario proposed to explain the constant spectral shape but strongly varying X-ray flux in NGC~2992 by \citet{Beckmann2007} and \citet{Middei2022}. In this model, the radio flux is attenuated by the passing coronal plasmoid, causing it to decrease, and the X-ray flux rises due to the increased number of up-scattered UV photons from the disk, without changing the X-ray spectrum appreciably. This magnetic reconnection scenario is described in detail by \citet{Gouveia2010}. In short, an increase in accretion rate creates strong ram pressure of inwardly-advected magnetic field lines against field lines anchored in the black hole's magnetosphere,  resulting in a region of very high $\beta$, the ratio of thermal and radiation pressure to magnetic pressure. Compressing lines of opposite polarization leads to violent reconnections. When a reconnection occurs, particles are accelerated and ejected at relativistic speeds and a ``luminous blob" is released. At first, the relativistic electrons in the blob are self-absorbed, but as the blob rises and expands adiabatically it becomes increasingly transparent to its own radiation and any radiation passing through to the observer. The varying flux but unchanging spectral shape (see Figure~\ref{fig:xray_spectra} of the X-ray observations presented here) are consistent with this picture; to an extent this can also be seen in the non-varying hardness ratio during the event in the bottom panel of Figure~\ref{fig:lcs}.

\subsubsection{Flare Energetics}
\label{sec:flare}

\citet{Gouveia2010} put forward a prescription for the amount of magnetic energy that can be extracted from a reconnection event based on black hole parameters, which they parametrize from stellar mass black holes all the way to luminous quasars:\\

\noindent\begin{equation}
\dot{W}_B = 1.6\times10^{35} \alpha_{0.5}^{-19/16} \beta_0.8^{-9/16} M_{14}^{19/32} R_{X,7}^{-25/32}l_{100}^{11/16} 
\end{equation} erg/s

Where $\alpha$ is the accretion disk viscosity parameter, $M_{14}$ is the black hole mass in units of $14M_\odot$, $R_X$ is the inner radius of the accretion disk (and $R_{X,7}$ is this number in units of $10^7$~cm), and $l_{100}$ is the scale height of the magnetic reconnection region in units of $100R_X$.  The black hole mass of UGC~6728 is well-constrained. The spin of the black hole is not known precisely but is estimated from \emph{Suzaku} spectral fitting to be on the high side, $a>0.7$ \citep{Walton2013}. For simplicity's sake, we set $R_X$ equal to the Schwarzschild radius, which is the approximate location of the innermost-stable circular orbit (and thus the accretion disk inner limit) in a maximally-spinning black hole. If we assume reasonable values of $\alpha = 0.2$ and $\beta=0.1$ (assuming the reconnection is only likely to happen when the magnetic pressure far exceeds the thermal and radiation pressure), we obtain $\dot{W}_B = 1.9\times10^{42}$~erg/s. 

To obtain the approximate amount of energy released during the X-ray flare, we integrate over the 20-day minimum flare duration using the composite Simpson's rule for irregularly spaced data, correcting the time axis to the galaxy's rest frame. The integrated number of counts in the flare is then converted to erg/cm$^2$ using the results of the XSPEC fit (1~ct = $3.89\times10^{-11}$~erg~cm$^2$) and then to total energy through multiplication by the luminosity distance.  The total energy released in the flare, assuming a 20-day duration, after subtracting the baseline flux throughout the duration based on the light curve points on either side, is $1.45\times10^{47}$~ergs. The instantaneous flux at the peak of the flare is 0.82~cts/s, which corresponds to a luminosity of $2.4\times10^{41}$~erg~s$^{-1}$. This is consistent with the estimated energy released in a typical X-ray flare due to magnetic disk activity of $\sim5\times10^{41}$~erg~s$^{-1}$ from \citet{DiMatteo1998}, and is approximately 10\% of the value calculated above for a magnetic event based on the parameters of UGC~6728.

\subsubsection{Properties of Obscuring Material}
\label{sec:blob_properties}

Here we follow the methodology of \citet{Fernandez2022} and calculate the optical depth required for the blob to cause the observed attenuation of the radio source, based on \citet{Osterbrock89}:

\begin{equation}
    \tau_\nu = 3.28\times10^{-7} \left(\frac{T}{10^4\mathrm{K}}\right)^{-1.35} \left(\frac{\nu}{\mathrm{GHz}}\right)^{-2.1} \left(\frac{\int n_+n_e ds}{\mathrm{pc~cm}^{-6}}\right) 
\end{equation}

We can assume that $n_+ = n_e$, reducing the integrand to $n_e^2$, and the full integral to $N_e$, the total column along the line of sight, $s$. The $5\sigma$ upper limit of the non-detection at the lowest point of the radio light curve is $6.4 \mu$Jy. To reduce the observed radio flux from the pre-dip point at 0.125~mJy to this upper limit, an optical depth of $\tau = \ln (F/F_0) = 2.97$ is required (the answer is $\tau = 1.97$ if the lowest \emph{detected} radio flux is used). This is higher than the optical depth required for the \citet{Fernandez2022} event, but they also used an upper limit, and their rms noise level is much higher than ours throughout their observations - so, it is possible that the flux got much lower. However, both yield reasonable physical conditions for the cloud.

\begin{figure}
\gridline{\fig{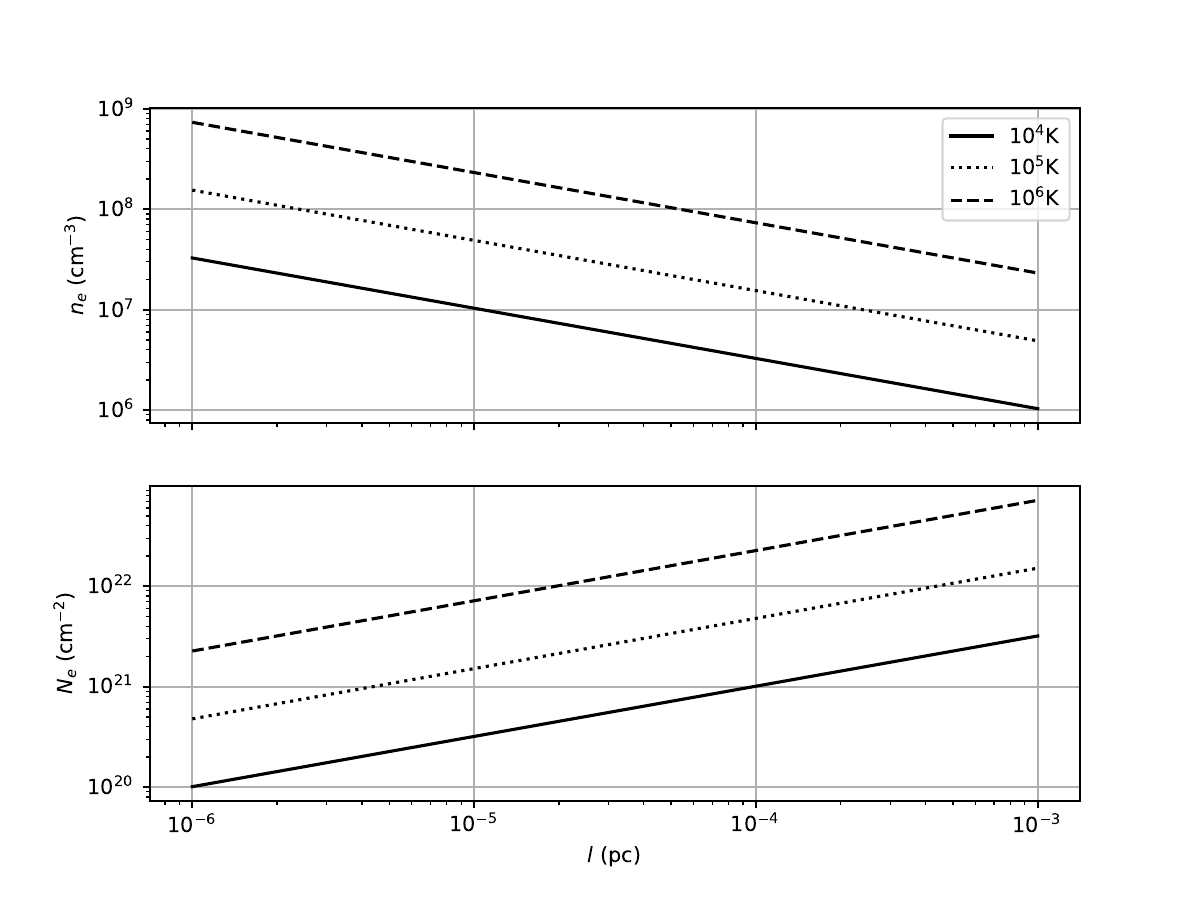}{0.5\textwidth}{}}
\caption{Dependence of electron density (top) and total electron column density (bottom) on the path length through the obscuring cloud for three representative temperatures.}
\label{fig:column_densities}
\end{figure}

We calculate the electron density $n_e$ as a function of the length of the path through the cloud, along with the total electron column $N_e$, for temperatures $T=10^4$, $10^5$, and $10^6~$K and plot the results in Figure~\ref{fig:column_densities}.

The low total electron column densities ($N_e$) are consistent with the X-ray spectral fitting results, showing negligible column. So, the X-ray spectral fits during the radio dip are not at odds with the properties of this particular obscuring cloud. As found by \citet{Fernandez2022}, the properties of the blob are broadly consistent with typical BLR conditions - however, the very weak $n_e \propto \sqrt{\tau}$ dependence means that the required optical depth can vary largely without requiring an order of magnitude increase in electron density. Nonetheless, the optical depth requirement for the observed dip of $\tau = 2.97$ is easily consistent with the expected conditions. 

\subsubsection{Occultation Timescale}
\label{sec:timescales}
In the light curves of UGC~6728, there is at least a 10-day delay between the flare diminishing and the radio flux beginning to decrease - this is the lower limit on the ``crossing time" between the blob ejection site and the moment it begins eclipsing the radio source. Since we do not have any radio flux information until 664 days after this, 674 days represents an upper limit on the plasmoid's travel time. However, by comparing the X-ray behavior with the event seen in NGC~2992, it is likely that the flux decreased gradually, beginning before MJD~60154. We note that the small redshift of UGC~6728, $z=0.0065$, introduces a time dilation correction of only a couple of days, which is negligible considering the uncertainties due to the light curve sampling. The round numbers estimated in the following sections are not affected significantly by this, so we neglect it, especially since the redshift of this target is somewhat uncertain (see earlier footnote).

To determine whether the blob occultation model is reasonable for the geometry of UGC~6728's central engine, it is necessary to assume a velocity of the blob as it travels upward away from the accretion disk. 

If the reconnection events described by \citet{Gouveia2010} are responsible for the putative eclipsing plasmoid in the case of UGC~6728 and NGC~2992, then they may be moving quite quickly - relativistically, in fact. Though such ultra-fast outflows (UFOs) do occur in some radio-quiet AGN \citep{Tombesi2010}, the X-ray spectrum of UGC~6728 shows no signs of a UFO or even a strong soft-excess. In archival X-ray spectra, \citet{Walton2013} found evidence of a slight soft excess, but it was consistent only with reflection and did not require an outflowing component. 

We could instead posit a plasmoid velocity in the range of observed narrow line shifts assumed to arise from radiatively-driven winds off the accretion disk - a few hundred km/s.

The black hole mass of UGC~6728 is well-constrained, measured by reverberation mapping to be $7.1\times10^5 M_\odot$ \citep{Bentz2016}. The Schwarzschild radius for such a mass is $R_S = 2GM/c^2 = 1.86\times10^{11}$cm, or about 0.012~AU. If we guess a blob velocity of 300~km/s and assume it travels straight upward away from the disk, it will travel 144~$R_S$ in 10 days and 9410~$R_S$ in 674 days. Both of these are far in excess of the estimated scale height of a ``lamp-post" style corona \citep{Alston2020}, which is on the order of a few to tens of $R_S$, even when it is found to depend upon the AGN luminosity \citep{Ursini2020,Wilkins2023,Laha2025}. The corona is also believed to be quite compact, less than 10$R_S$ in size, based on rapid X-ray variability \citep{Fabian2015}. Therefore, the radio source being obscured by the rising blob cannot be the AGN corona if it is assumed to be a lamp-post style model. Even in newer models, where the corona is elongated or outflowing, its size is restricted to a few tens of $R_S$ \citep[e.g., ][]{Zhao2025}. Since \citet{Fernandez2022} did not have X-ray monitoring preceding the event, we cannot calculate a flare-to-eclipse timescale for NGC~2992. However, in that galaxy's case, the probability of the eclipsed source being the corona is more reasonable: if we assume that the event begins when the X-ray flux starts to rise, and lasts 140 days, and assuming the same 300~km/s speed for the blob, the implied coronal size is $\sim12R_S$, consistent with the studies cited here. This is due both to the much shorter event duration and the much larger black hole mass of $10^8 M_\odot$ (although this mass is not nearly as well constrained as that of UGC~6728). If the same mechanism is at play here, either the occulted radio source must be much larger than the corona (e.g., a nascent, low-powered jet) or the obscuring blob must be much larger and rapidly expanding, enabling it to obscure the radio source for a longer time.

There are several options for a larger radio source: the origin of radio emission in radio quiet AGN is likely to be composite, including the corona and interactions between an outflowing wind or jet \citep{Panessa2019}. While it is broadly accepted that some fraction of radio emission in radio-quiet AGN comes from the corona \citep[e.g., ][]{Laor2008}, there is increasing evidence that some, or even most, of this emission comes from larger scales, perhaps within the broad line region due to a wind/gas interaction \citep{Fischer2021,Chen2024}. However, it is not clear how such a large source would be obscured by a blob released in the inner disk. So, if obscuration of the radio source is still to be viable, the radio source must be a narrow, collimated one, like a jet. If, despite the poorly sampled light curve, the obscuration is indeed a single event, its full duration is 1171~days, meaning that the radio source was at least partly obscured over 16290$R_S$, or $9.8\times10^{-4}$~pc ($\sim203$~AU). This is still well inside the unresolved 0.2\arcsec~beam size of our observations, which spatially corresponds to 270~parsecs. The source was not detected in a shallow VLBA fringe survey by \citet{Baek2019}, but their detection limit was 13~mJy, well above the baseline radio flux in UGC~6728. So, we have no information about the radio structure on smaller scales. We can gain some insight from the spectrum at this point: there are simultaneous C- and X-band observations at MJD~60154 and 60161. The C-band does not detect a source in either observation, but the source \emph{is} detected in the first X-band observation, well above the 3-$\sigma$ upper limit of the C-band. This means that the radio spectrum was inverted: detected robustly at $\sim9$~GHz and much fainter at $\sim$5~GHz. An inverted radio spectrum is an indication of a young or newly-launched jet \citep[e.g., ][]{O'Dea2021,Cheng2023}, which is potentially consistent with a small but elongated radio source.

\subsection{Coronal Flaring}

The X-ray behavior of UGC~6728 is not unusual taken by itself. Flaring behavior on the order of a factor of a few over timescales of minutes to days is a normal occurrence in the X-ray light curves of radio-quiet AGN and has been known for decades \citep[e.g., ][and many others]{Lawrence1987,McHardy2005,Gonzalez2012,Papadakis2024}. Indeed, this rapid variability is key in constraining the size of the coronal region, which is limited in some cases to as compact as 10 Schwarzschild radii \citep{Laha2025}. So, it is possible that the flare seen in the X-ray light curve presented here is just normal coronal variability, the continuance of which is not observed due to the subsequent sparseness of the X-ray cadence. The similarly sparse radio light curve may, as mentioned at the beginning of this section, also mask standard variability, although it is not at all certain that the radio and X-ray light curves vary together even in radio-quiet AGN without indications of a jet (see discussion in the introduction). 

During coronal flaring, however, some spectral changes are expected for most models. \citet{Wilkins2014} and \citet{Gallo2019} report a softening during the brightening phenomenon. This is interpreted by \citet{Wilkins2014} as evidence that the corona expands with increasing luminosity (i.e., as more energy is introduced into the plasma by the accretion flow) and contracts with decreasing luminosity, and by \citet{Gallo2019} as supporting the corona as the base of a radio jet, which extends vertically in high-luminosity states. \citet{Ding2022} report changing compactness and coronal temperature in the super-Eddington radio-intermediate AGN IZW~1, which may have triggered a pair production runaway causing rapid cooling. Because the X-ray spectrum below 10~keV is comprised of both a hard power law due to the comptonization of seed photons from the disk and a soft excess which may come from relativistically blurred reflection off the disk \citep[e.g., ][ and references therein]{Ross2005,Crummy2006, Laha2025}, flares due to increased accretion rate or coronal geometry are likely to result in spectral changes between high- and low-flux states. The flare observed in the light curve presented here in UGC~6728 shows no X-ray spectral variation at all compared to other epochs (Figure~\ref{fig:xray_spectra}), which may indicate that a typical coronal flare is not at play here. A magnetic reconnection in the \emph{accretion disk}, however, may not influence the X-ray spectrum significantly, as predicted by the model of \citet{Gouveia2010}(Section~\ref{sec:magrecon}. Such reconnections can occur far from the truly nuclear corona, either due to normal magnetic turbulence from differential rotation at many disk radii \citep[e.g., ][]{DiMatteo1998} or more exotic scenarios like inward migration of satellite black holes \citep{Xing2025}.

\section{Conclusions} 
\label{sec:conclusion}
We report the observation of a possible temporary obscuration event of the radio source in the dwarf AGN UGC~6728, in which the X-ray emission flares, returns to the pre-flare value, and then rises slowly over 1171 days before decreasing back to its baseline. At the same time as this event, the radio flux decreases to non-detection, and then rises back to its steady value before the X-ray flare. This behavior is very similar to an event seen in NGC~2992 by \citet{Fernandez2022} and theorized to be a blob emitted by magnetic reconnection in the inner accretion disk, which rises and expands, increasing the X-ray flux, and temporarily partially obscures the nuclear radio source. In that object, no pre-event flare was seen, as we see in UGC~6728. We have investigated the X-ray flare, X-ray spectrum, and necessary conditions of the obscuring medium to produce the observed dimming of the radio flux. We come to the following conclusions:

\begin{itemize}
    \item The X-ray flare seen before the possible obscuration event reaches a peak luminosity of $2.4\times10^{41}$~erg~s$^{-1}$, consistent with magnetic disk activity. This potentially supports of the interpretation of this and the NGC~2992 event as obscuration of the radio source by a plasmoid released by magnetic reconnection in the inner accretion disk.

    \item The X-ray spectrum and hardness do not vary significantly throughout the event, just as seen in NGC~2992, consistent with the idea that the X-ray flux is rising due to an increase in the total number of up-scattered UV seed photons in the corona, which would not alter its spectral shape. 

    \item The electron density and column density required to account for the observed dip in radio emission, $\tau = 2.97$, are consistent with expected conditions in the broad line region.

    \item The duration of the event is the strongest constraint on the proposed model. Unlike the event in NGC~2992, the obscured radio source cannot be the corona, because the 1171-day duration, assuming a blob velocity typical of AGN outflows, implies a source size of approximately $16000$ gravitational radii. Therefore, if occultation is the explanation for the radio dimming, it must be partially obscuring a more elongated source, such as a small-scale jet or extended radio emission in an outflow.
\end{itemize}

 While the light curve is by no means well-sampled, it has the same number of data points as the NGC~2992 event. It includes an X-ray flare preceding the possible obscuration, as would be expected from the \citet{Fernandez2022} model for a reconnection-launched blob as the obscurer. If such events are observed more regularly as radio time domain surveys become feasible, such as with the upcoming DSA-2000 array \citep{Hallinan2019}, they may provide an observable handle on the size of radio sources and the population statistics of plasmoids released by magnetic reconnections, and thereby, the properties of AGN inner accretion disks. X-ray flares such as those seen before the eclipsing event could also be considered as triggers for target-of-opportunity radio timing, to better track the evolution of the density and velocity of the obscuring material and map the causal relationships of the AGN central engine.

\begin{acknowledgments}
KLS gratefully acknowledges support from the Mitchell-Heep-Munnerlyn Career Enhancement chair at Texas A\&M. This work was supported in part by NASA grant 80NSSC24K0500. 
\end{acknowledgments}





%
\facilities{Swift-XRT, JVLA}

\software{XSPEC \citep{Arnaud1996}, CASA \citep{Casa2022}, CARTA \citep{Comrie2024}}


\bibliography{reference}{}
\bibliographystyle{aasjournalv7}



\end{document}